\documentclass[a4paper,fleqn,usenatbib]{mnras}

\usepackage[T1]{fontenc}
\usepackage{ae,aecompl}

\usepackage{graphicx}	
\usepackage{amsmath}	
\usepackage{amssymb}	
\usepackage{graphics}

\usepackage{xcolor}

\title[HTRU - XIII. The most accelerated binary pulsar]{The High Time Resolution Universe Pulsar Survey - XIII. PSR~J1757$-$1854, the most accelerated binary pulsar}

\author[A. D. Cameron et al.]{
A.~D.~Cameron$^{1}$\thanks{E-mail: acameron@mpifr-bonn.mpg.de},
D.~J.~Champion$^{1}$,
M.~Kramer$^{1,2}$,
M.~Bailes$^{3,4,5}$,
E.~D.~Barr$^{1}$,
\newauthor
C.~G.~Bassa$^{6}$,
S.~Bhandari$^{3,4}$,
N.~D.~R.~Bhat$^{4,7}$,
M.~Burgay$^{8}$,
S.~Burke-Spolaor$^{9,10}$,
\newauthor
R.~P.~Eatough$^{1}$,
C.~M.~L.~Flynn$^{3}$,
P.~C.~C.~Freire$^{1}$,
A.~Jameson$^{3,4}$,
S.~Johnston$^{11}$,
\newauthor
R.~Karuppusamy$^{1}$,
M.~J.~Keith$^{2}$,
L.~Levin$^{2}$,
D~.R.~Lorimer$^{9}$,
A.~G.~Lyne$^{2}$,
\newauthor
M.~A.~McLaughlin$^{9}$,
C.~Ng$^{12}$,
E.~Petroff$^{6}$,
A.~Possenti$^{8}$,
A.~Ridolfi$^{1}$,
B.~W.~Stappers$^{2}$,
\newauthor
W.~van~Straten$^{3,4,13}$,
T.~M.~Tauris$^{1,14}$,
C.~Tiburzi$^{1,15}$,
N.~Wex$^{1}$
\\
$^{1}$Max-Planck-Institut f{\"u}r Radioastronomie, Auf dem H{\"u}gel 69, D-53121 Bonn, Germany.\\
$^{2}$Jodrell Bank Center for Astrophysics, University of Manchester, Alan Turing Building, Oxford Road, Manchester M13 9PL, United Kingdom.\\
$^{3}$Centre for Astrophysics and Supercomputing, Swinburne University of Technology, Mail H39, PO Box 218, VIC 3122, Australia.\\
$^{4}$ARC Center of Excellence for All-Sky Astronomy (CAASTRO), Swinburne University of Technology, Mail H30, PO Box 218, VIC 3122, Australia.\\
$^{5}$ARC Center of Excellence for Gravitational Wave Discovery (OzGrav), Swinburne University of Technology, Mail H11, PO Box 218, VIC 3122, Australia.\\
$^{6}$ ASTRON, the Netherlands Institute for Radio Astronomy, Postbus 2, NL-7990 AA Dwingeloo, the Netherlands.\\
$^{7}$International Centre for Radio Astronomy Research, Curtin University, Bentley, WA 6102, Australia.\\
$^{8}$INAF - Osservatorio Astronomico di Cagliari, Via della Scienza 5, I-09047 Selargius (CA), Italy.\\
$^{9}$Department of Physics and Astronomy, West Virginia University, PO Box 6315, Morgantown, WV 26506, USA.\\
$^{10}$Center for Gravitational Waves and Cosmology, West Virginia University, Chestnut Ridge Research Building, Morgantown, WV 26505, USA.\\
$^{11}$CSIRO Astronomy \& Space Science, Australia Telescope National Facility, P.O. Box 76, Epping, NSW 1710, Australia.\\
$^{12}$Department of Physics and Astronomy, University of British Columbia, 6224 Agricultural Road, Vancouver, BC V6T 1Z1, Canada.\\
$^{13}$Institute for Radio Astronomy \& Space Research, Auckland University of Technology, Private Bag 92006, Auckland 1142, New Zealand.\\
$^{14}$Argelander-Insitut f{\"u}r Astronomie, Universit{\"a}t Bonn, Auf dem H{\"u}gel 71, 53121 Bonn, Germany\\
$^{15}$Fakult{\"a}t f{\"u}r Physik, Universit{\"a}t Bielefeld, Postfach 100131, D-33501 Bielefeld, Germany.\\
}

\date{Accepted XXX. Received YYY; in original form ZZZ}

\pubyear{2017}

\begin{document}
\label{firstpage}
\pagerange{\pageref{firstpage}--\pageref{lastpage}}
\maketitle

\begin{abstract}
We report the discovery of PSR~J1757$-$1854, a $21.5$-ms pulsar in a highly-eccentric, $4.4$-h orbit with a neutron star (NS) companion. PSR~J1757$-$1854 exhibits some of the most extreme relativistic parameters of any known pulsar, including the strongest relativistic effects due to gravitational-wave (GW) damping, with a merger time of 76\,Myr. Following a 1.6-yr timing campaign, we have measured five post-Keplerian (PK) parameters, yielding the two component masses ($m_\text{p}=1.3384(9)\,\text{M}_\odot$ and $m_\text{c}=1.3946(9)\,\text{M}_\odot$) plus three tests of general relativity (GR), which the theory passes. The larger mass of the NS companion provides important clues regarding the binary formation of PSR~J1757$-$1854. With simulations suggesting 3-$\sigma$ measurements of both the contribution of Lense-Thirring precession to the rate of change of the semi-major axis and the relativistic deformation of the orbit within $\sim7-9$ years, PSR~J1757$-$1854 stands out as a unique laboratory for new tests of gravitational theories.

\end{abstract}




\section{Introduction}\label{sec: Introduction}

Since the discovery of the Hulse-Taylor pulsar \citep[PSR~B1913$+$16;][]{ht75}, binary pulsars have played a critical role in providing key tests of GR and its alternatives. Even with the direct detection of gravitational waves \citep{aaa+16} and the more recent direct observation of a double neutron star (DNS) merger \citep{aaa+17}, binary pulsars are still able to provide some of the most important gravity tests with strongly self-gravitating bodies, particularly in the quasi-stationary strong-field gravity regime \citep{wex14}. An example is the constraint on the leading-order GW emission in GR (as described by the quadrupole formula), for which the double pulsar \citep[PSR~J0737$-$3039;][]{bdp+03, lbk+04} currently stands out as the most constraining system, surpassing LIGO by three orders of magnitude \citep{kra16, aaa+16}. PSR~J0737$-$3039 currently offers five independent tests of GR (based on six PK parameters and the mass ratio), of which it passes the most stringent to within a measurement uncertainty of only $0.05\%$ \citep{ksm06x,bkk+08}. Other binary pulsars, such as the millisecond pulsar-white dwarf PSR~J1738$+$0333, provide strong constraints on dipolar GW emission, a prediction of many alternative theories of gravity such as scalar-tensor theories \citep{fwef+12}.

Pulsar constraints on the nature of GWs, the limits of GR and on alternative gravitational theories are anticipated to provide both complementary and competitive tests against the most advanced ground-based GW detectors currently foreseen \citep{ssb+17}. However, this depends upon the discovery of additional relativistic systems capable of expanding the parameter space currently explored by binary pulsars. To this end, the High Time Resolution Universe South Low Latitude pulsar survey \citep[HTRU-S LowLat,][]{kjvs10, ncb15}, conducted using the Parkes 64-m radio telescope, was undertaken with the specific goal of discovering additional relativistic binary pulsars. This survey covers the inner Galactic plane ($-80^\circ < l < 30^\circ$ and $\left|b\right|<3.5^\circ$) and is the region predicted to contain the highest number of relativistic binaries \citep{bkb02}.

\begin{figure}
\begin{center}
\includegraphics[height=\columnwidth, angle=270]{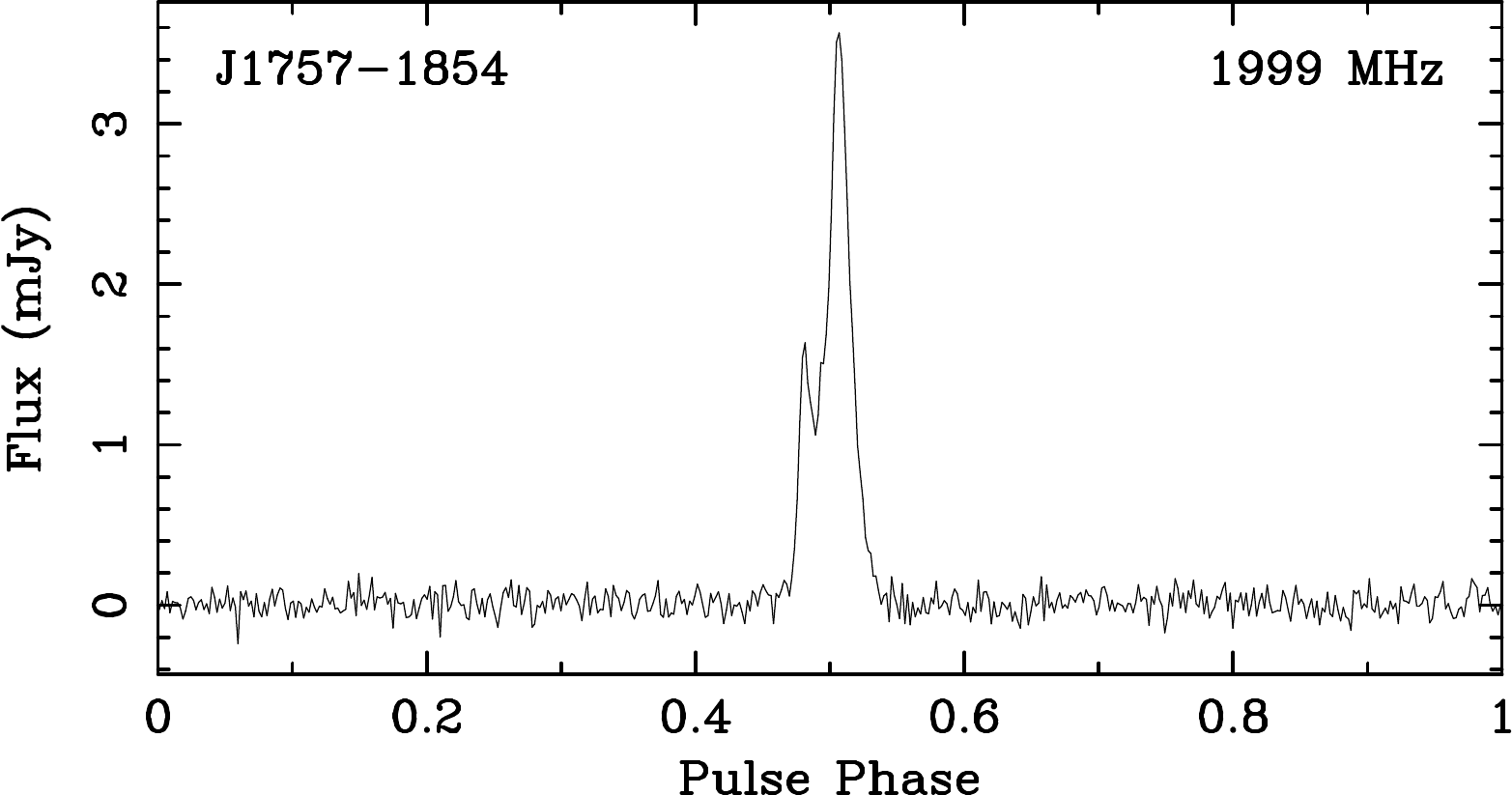}
\end{center}
\caption{Pulse profile of PSR~J1757$-$1854, observed with the Green Bank Telescope (GBT) on MJD~57857, integrated over approximately one full orbit.}\label{fig: profile}
\end{figure}

Here, we report the discovery of PSR~J1757$-$1854, the first relativistic binary discovered in the HTRU-S LowLat survey. PSR~J1757$-$1854 (see Fig. \ref{fig: profile}) is a $21.5$-ms pulsar in a $4.4$-h orbit with an eccentricity of 0.61 and a NS companion, making the system a DNS. The compactness, high eccentricity and short orbital period of PSR~J1757$-$1854 make it one of the most relativistic binary pulsars known, with the potential for even more rigorous constraints to be placed on GR and other gravitational theories.


\section{Discovery}\label{sec: Discovery}

\subsection{Candidate identification and confirmation}\label{subsec: identification}

To search for binary pulsars, we employ the `time-domain resampling' technique \citep[see e.g.][]{mk84,jk91}, which assumes that the binary motion can be modelled as a constant line-of-sight acceleration. For a circular orbit this assumption holds best when the quantity $r_\text{b} = t_\text{int}/P_\text{b}\leq0.1$ where $t_\text{int}$ is the integration time of the observation and $P_\text{b}$ is the orbital period \citep[see e.g.][]{jk91,ncb15}. Building on the technique of \cite{eklk+13}, our `partially-coherent segmented acceleration search' uses this principle to blindly search for pulsars in compact binary systems by progressively halving each observation into smaller time segments (as low as $t_\text{int}=537\,\text{s}$) which are independently searched (to accelerations as high as $\left|a\right|=1200\,\text{m\,s}^{-2}$). This has the trade-off of increasing our sensitivity to shorter orbital periods while gradually reducing our sensitivity in flux density. Full details of the technique are available in \cite{ncb15}.

PSR~J1757$-$1854 was identified in the second 36-min half-length segment of a full 72-min observation recorded on MJD 56029, with an acceleration of $-32\,\text{m\,s}^{-2}$ and a signal-to-noise ratio ($\text{S/N}$) of 13.3. The pulsar's signal was recoverable across the full observation to a S/N of 21.4, but with a significantly-changing acceleration (i.e. `jerk'). Consequently, a time-domain acceleration search of the full-length observation only detected the pulsar at a reduced S/N of only 10.6, indicating that the segmented search greatly assisted in the discovery of this pulsar. At periastron, the pulsar reaches a maximum absolute acceleration of $\sim684\,\text{m\,s}^{-2}$, the highest of any known binary pulsar system. 

\subsection{Non-detection in the PMPS}\label{subsec: PMPS}

The Parkes Multibeam Pulsar Survey \citep[PMPS,][]{mlc+01} contains a beam coincident with the timed position of PSR~J1757$-$1854 (see Table \ref{tab: timing solution}). However, neither acceleration searches using both time-domain resampling and matched filtering \citep{rem02} nor a fold using the ephemeris in Table \ref{tab: timing solution} detected the pulsar in the PMPS data. Based upon the radiometer equation \citep[see e.g.][]{lk05} and the dispersive smearing introduced by the 3-MHz channel size of the PMPS, we expect a $\text{S/N}_\text{PMPS}\simeq8$, rendering any detection of PSR~J1757$-$1854 difficult. Additional factors such as geodetic precession may also play a role, but at present a precise cause cannot be determined.


\section{Timing}

\subsection{Observations and data reduction}

A summary of PSR~J1757$-$1854's timing observations can be found in Table \ref{tab: timing config}. Parkes (PKS) observations used the 21-cm Multibeam \citep[MB20;][]{swb+96} and H-OH receivers, in combination with the Berkeley Parkes Swinburne Recorder\footnote{https://astronomy.swin.edu.au/pulsar/?topic=bpsr} (BPSR), the CASPER Parkes Swinburne Recorder\footnote{https://astronomy.swin.edu.au/pulsar/?topic=caspsr} (CASPSR) and a Digital Filter Bank system (DFB4). Jodrell Bank (JBO) observations with the 76-m Lovell telescope employed an L-Band receiver with a ROACH backend system \citep{bjk+16}. Observations at Effelsberg (EFF) were performed with the 7-Beam receiver with the PSRIX backend \citep{psrix} operating in both a folded and baseband recording mode (FOLD and BB respectively). Finally, observations with the Green Bank Telescope (GBT) were conducted using the L-Band, S-Band and Prime-Focus 800-MHz (PF1-800) receivers, all in combination with the Green Bank Ultimate Pulsar Processing Instrument \citep[GUPPI;][]{guppi}. All GBT and Effelsberg observations were designed to sample a full or significant fraction of the orbit.

\begin{table}
\caption{Telescope frontend and backend configurations, including the central frequency ($f_\text{c}$), bandwidth ($\Delta f$) of each combination and the number ($n_\text{TOA}$) and time span of the TOAs.}\label{tab: timing config}
\begin{center}
\begin{tabular}{ccrcc}
\hline
Receiver & Backend & \multicolumn{1}{c}{$f_\text{c}\left(\Delta f\right)$} & $n_\text{TOA}$ & Span\\
 &  & \multicolumn{1}{c}{(MHz)} & & (MJD) \\
\hline
\multicolumn{5}{l}{\textit{PKS:}}\\
MB20 & BPSR  & 1382(400) & 9 & 57405$-$57406 \\
& CASPSR$^{\text{a}}$ & 1382(400) & 41 & 57734$-$57986 \\
H-OH & DFB4 & 1369(256) & 57 & 57553$-$57675 \\
 & CASPSR$^{\text{a}}$ & 1382(400) & 55 & 57596$-$57635 \\
\hline
\multicolumn{5}{l}{\textit{JBO:}}\\
L-Band & ROACH$^{\text{b}}$ & 1527(400) & 422 & 57456$-$57958 \\
\hline
\multicolumn{5}{l}{\textit{EFF:}}\\
7-Beam & FOLD & 1360(240) & 83 & 57573$-$57896 \\ 
 & BB$^{\text{a}}$ & 1360(240) & 84 & 57815$-$57986 \\
\hline
\multicolumn{5}{l}{\textit{GBT:}}\\
PF1-800 & GUPPI$^{\text{a\phantom{,b}}}$ & 820(200) & 25 & 57620$-$57621 \\
L-Band & GUPPI$^{\text{a,b}}$ & 1499(800) & 731 & 57795$-$57950 \\
S-Band & GUPPI$^{\text{a,b}}$ & 1999(800) & 1655 & 57627$-$57998 \\
\hline
\multicolumn{5}{l}{$^{\text{a}}$ Observations recorded with coherent de-dispersion.} \\
\multicolumn{5}{l}{$^{\text{b}}$ $\Delta f$ split into 200 MHz sub-bands before TOA production.} \\
\end{tabular}
\end{center}
\end{table}

Data reduction employed the \textsc{dspsr} \citep{dspsr}, \textsc{psrchive} \citep{psrchive}, \textsc{sigproc}\footnote{http://sigproc.sourceforge.net}, \textsc{presto} \citep{ransom01}, \textsc{tempo}\footnote{http://tempo.sourceforge.net} and \textsc{tempo2} \citep{tempo2} software packages. Each TOA set was produced using its own reference profile (all rotated to the same pulse phase), and was weighted such that its reduced $\chi^2=1$. The data sets were combined using jumps fit across regions of common overlap. During this process, the pulsar's dispersion measure (DM) appeared to vary as a function of orbital phase. This resulted from an apparent inability of \textsc{dspsr} and \textsc{psrchive} to correctly de-disperse across a large bandwidth during rapid orbital motion, despite this effect having been accounted for by \textsc{tempo2}'s phase predictors \citep{tempo2}. To counteract this, Jodrell Bank and GBT (L and S-Band) TOAs were produced from $200$-MHz sub-bands.


\subsection{Measured parameters and implications}
\label{sec:params}

\begin{table}
\caption{Ephemeris of PSR~J1757$-$1854, as derived using \textsc{tempo2}. Numbers in parentheses represent 1-$\sigma$ uncertainties, with TOA errors re-weighted such that the reduced $\chi^{2}$ went from $1.7$ to $1.0$. DM distances are derived from the NE2001 \citep{NE2001a} and YMW16 \citep{YMW16} models.}
 \begin{center}\label{tab: timing solution}
  \begin{tabular}{lc}
   \hline
   Right ascension, $\alpha$ (J2000) & 17:57:03.78438(6) \\
   Declination, $\delta$ (J2000) & $-$18:54:03.376(7) \\
   Spin period, $P$ (ms) & 21.497231890027(7) \\
   Spin period derivative, $\dot{P}$ ($10^{-18}$) & 2.6303(7) \\
   Timing epoch (MJD) & 57701\\
   Dispersion measure, DM (pc\,cm$^{-3}$) & 378.203(2) \\
   \\
   Binary model & DDH \\
   Orbital period, $P_\text{b}$ (d) & 0.18353783587(5) \\
   Eccentricity, $e$ & 0.6058142(10) \\
   Projected semimajor axis, $x$ (lt-s) & 2.237805(5) \\
   Epoch of periastron, $T_0$ (MJD) & 57700.92599420(5) \\
   Longitude of periastron, $\omega$ ($^\circ$) & 279.3409(4) \\
   Rate of periastron advance, $\dot{\omega}$ ($^\circ\,\text{yr}^{-1}$) & 10.3651(2) \\
   Einstein delay, $\gamma$ (ms) & 3.587(12) \\
   Orbital period derivative, $\dot{P_\text{b}}$ ($10^{-12}$) & $-$5.3(2) \\
   Orthometric amplitude, $h_{3}$ ($\mu\text{s}$) & 4.6(7) \\
   Orthometric ratio, $\varsigma$ & 0.90(3) \\ 
   \\
   Mass function, $f$ ($\text{M}_\odot$) & 0.35718891(2) \\
   Total system mass, $M$ ($\text{M}_\odot$) & 2.73295(9)$^{\dagger}$ \\
   Pulsar mass, $m_\text{p}$ ($\text{M}_\odot$) & 1.3384(9)$^{\dagger}$ \\
   Companion mass, $m_\text{c}$ ($\text{M}_\odot$) & 1.3946(9)$^{\dagger}$ \\
   Inclination angle, $i$ ($^\circ$) & $84.0^{+0.4}_{-0.3}\text{ or }96.0^{+0.3}_{-0.4}$$^{\dagger}$ \\
   \\
   Flux density at 1.4 GHz, $S_{1400}$ (mJy) & 0.25(4) \\
   DM distance, $d$ (kpc) & 7.4 (NE2001) \\
    & 19.6 (YMW16) \\
   Surface magnetic field, $B_\text{surf}$ ($10^{9}\,\text{G}$) & 7.61 \\
   Characteristic age, $\tau_\text{c}$ (Myr) & 130 \\
   Spin-down luminosity, $\dot{E}$ ($10^{30}\,\text{ergs s}^{-1}$) & 10500 \\
   \\
	Time units & TCB \\
    Solar system ephemeris & DE421\\
    RMS residual ($\mu \text{s}$)& 36 \\
	\hline
    \multicolumn{2}{l}{$^{\dagger}$ Parameters derived according to the DDGR model.}
  \end{tabular}
 \end{center}
\end{table}

Our derived ephemeris of PSR~J1757$-$1854, employing the DDH \citep{fw10} binary model, is provided in Table \ref{tab: timing solution}. Based upon the spin parameters, we derive a characteristic age $\tau_\text{c}\simeq130\,\text{Myr}$ and a surface magnetic field $B_\text{surf}\simeq7.61\times10^{9}\,\text{G}$, indicating that the pulsar has been partially recycled. Five PK parameters, including the rate of periastron advance $\dot{\omega}$, Einstein delay $\gamma$, orbital period derivative $\dot{P}_\text{b}$ and orthometric Shapiro parameters $h_{3}$ and $\varsigma$, have been measured significantly. Using the DDGR model \citep{taylor87,tw89}, which assumes the correctness of GR, we derive the total system mass $M=2.73295(9)\,\text{M}_\odot$ and the separate masses of the pulsar ($m_\text{p}=1.3384(9)\,\text{M}_\odot$) and its companion ($m_\text{c}=1.3946(9)\,\text{M}_\odot$). These masses, along with the high eccentricity and an implied second supernova (see Section \ref{sec:evolution}) indicate that the system is a DNS. From $m_\text{p}$, $m_\text{c}$ and the mass function we can further infer an inclination angle of $i=84.0^{+0.4}_{-0.3}\,^{\circ}$ (or $96.0^{+0.3}_{-0.4}\,^{\circ}$, when accounting for the $i \leftrightarrow 180^\circ - i$ ambiguity of the mass function), i.e., the orbit appears to be close to edge-on.
 
\begin{figure}
\begin{center}
\includegraphics[width=\columnwidth]{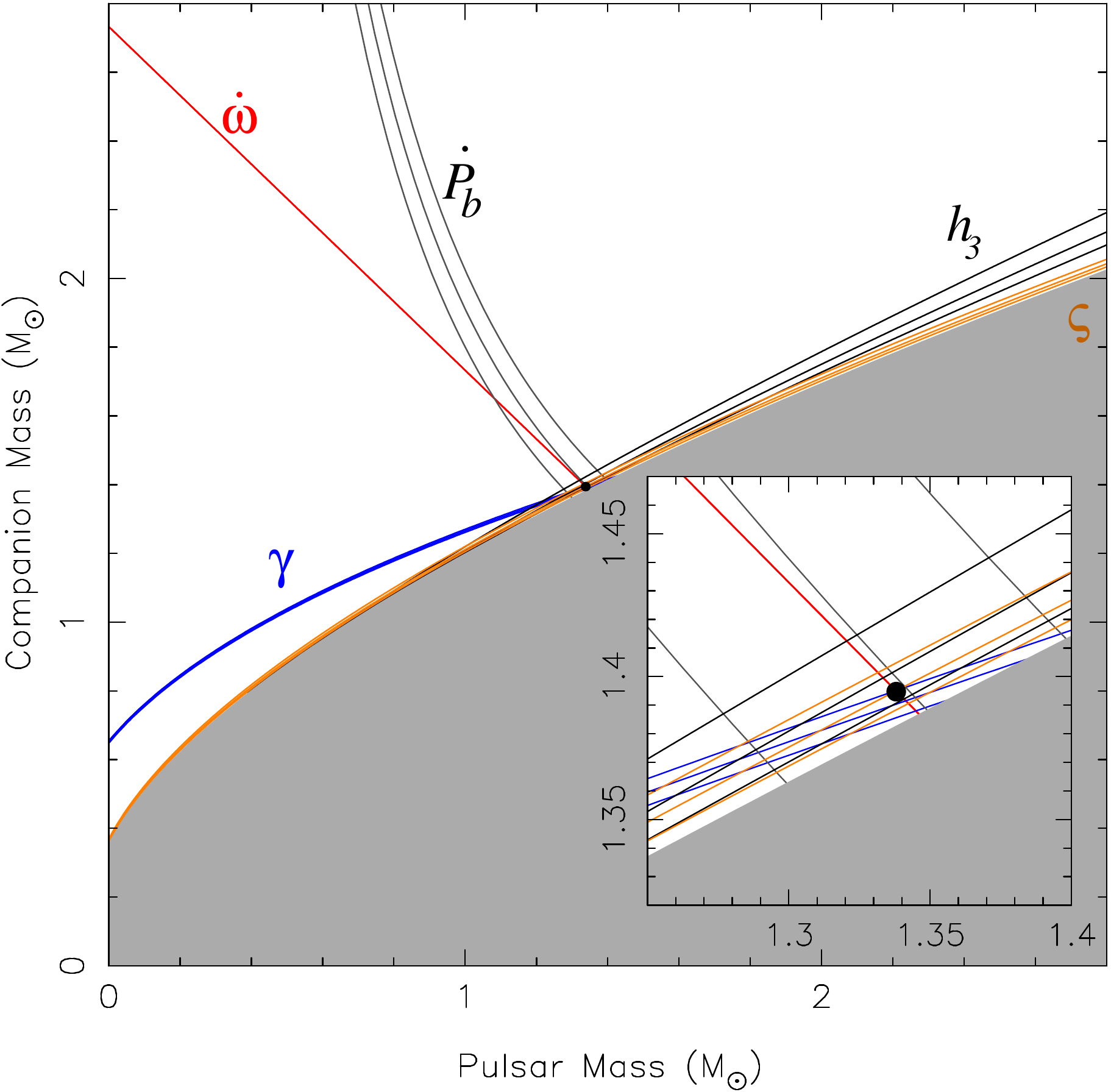}
\end{center}
\caption{Mass-mass diagram for PSR~J1757$-$1854. Shown are the mass constraints imposed under GR by each PK parameter, along with their 1-$\sigma$ error bars. A zoomed view of the region of intersection is shown in the inset, with the black dot indicating the DDGR masses. The grey region is excluded by orbital geometry.}\label{fig: mass mass}
\end{figure}

Fig. \ref{fig: mass mass} shows the constraints on the NS masses derived from the measured PK parameters under the assumption of GR. By using the intersection of $\dot{\omega}$ and $\gamma$ to fix the two NS masses, we can derive three new tests of GR from the remaining PK parameters. Based upon the $\dot{\omega}$-$\gamma$ mass solution, GR predicts an orbital decay due to GW damping of $\dot{P}_\text{b}=-5.2747(6)\times10^{-12}$, which the measured value of $\dot{P}_\text{b}$ agrees with to within a relative uncertainty of only 5\%. For $\varsigma$ and $h_3$, both observed values are within 1-$\sigma$ agreement of their GR predicted values ($0.92^{+0.040}_{-0.025}$ and $5.37^{+0.72}_{-0.40}\,\mu\text{s}$ respectively), indicating that GR passes all three tests.

PSR~J1757$-$1854 exceeds many of the relativistic qualities of previous binary pulsars, setting records (among others) for the closest binary separation at periastron ($0.749\,\text{R}_\odot$) and the highest relative velocity ($1060\,\text{km\,s}^{-1}$) at periastron. It also shows the strongest effects of GW damping yet seen in a relativistic pulsar binary, displaying the highest value of $\dot{P}_\text{b}$ as well as the highest value of $\dot{P}_\text{b}/P_\text{b} = -3.33 \times 10^{-16}\,\text{s}^{-1}$, the leading-order term in the cumulative shift in periastron time \citep[see e.g.][]{tw82}. This results in an inferred merger time of $76\,\text{Myr}$. Hence, PSR~J1757$-$1854 can be seen to probe a relativistic parameter space not yet explored by previous binary pulsars.

\subsection{Searches for the companion NS}
\label{subsec: companion search}

All GBT observations were recorded in coherently de-dispersed search mode, with two methods used to search for pulsations from the companion NS. The first method involved breaking each de-dispersed time series into $\sim30$-min segments, before performing a matched-filter acceleration search using the \textsc{presto} program \textsc{accelsearch} with a filter width of $z_\text{max}=50$ (where $z$ is the number of Fourier bins drifted by an accelerating pulsar). The second method (described in \citealt{msf+15}) involved resampling each time series in order to deconvolve the orbital motion of the companion NS as anticipated from the measured values of $m_\text{c}$ and $m_\text{p}$, before applying a \textsc{presto}-based periodicity search. Neither method detected pulsations from the companion. As precession may move the beam of the companion into the line-of-sight in the future, we will continue to record search-mode data so as to repeat these studies regularly.


\section{Evolutionary history}
\label{sec:evolution}

PSR~J1757$-$1854 is remarkable given its combination of a short orbital period, a large eccentricity, a relatively massive young NS companion (more massive than the recycled pulsar, a property shared with only one other published DNS system, PSR~B1534+12, \citealt{sttw02}), and a fast spinning recycled pulsar. The fast spin is expected for such a short orbital period DNS system, since in tight systems the recycling of the first-born NS is most efficient \citep[see Case~BB mass transfer modelling in][]{tlp15,tkf+17}.

Both the large eccentricity and the relatively massive young NS hint that a large kick is likely to have been imparted on the young NS at birth; see e.g. the mass--kick correlation suggested by \citet{tkf+17}. Indeed the two DNS systems B1913+16 and B1534+12, which also have relatively massive young NS companions, were shown to have experienced kicks of $\sim 200-400\,\text{km\,s}^{-1}$. 

To test this hypothesis for PSR~J1757$-$1854, we performed Monte Carlo simulations of the kinematic effects of the second supernova following the method outlined in \citet{tkf+17}. As expected, we find that a large kick is most likely at work for this system (the broad distribution of solutions peaks at a value near $400\,\text{km\,s}^{-1}$). For the mass of the exploding star, we find solutions from less than $2\,\text{M}_{\odot}$ and up to our maximum input limit of $7\,\text{M}_{\odot}$. However, the distribution peaks at the smallest value, supporting the idea of an ultra-stripped star exploding \citep{tlm+13,tlp15}.


\section{Future prospects}
\label{sec: future prospects}

The simulations described in Section \ref{sec:evolution} also produce a systemic 3D velocity distribution for PSR~J1757$-$1854, which peaks at a value of the order $200\,\text{km\,s}^{-1}$. Assuming a representative velocity in the plane of the sky of $150\,\text{km\,s}^{-1}$ and the NE2001 DM distance of 7.4-kpc (see Table \ref{tab: timing solution}) produces a predicted proper motion of $4.3\,\text{mas\,yr}^{-1}$. Furthermore, for the recycled pulsar we also obtain a distribution of misalignment angles between the spin vector of the pulsar and the orbital angular momentum, with a median value of $\sim25^\circ$. 

Consequentially, PSR~J1757$-$1854 is also expected to allow for future measurements of Lense-Thirring precession. Due to the large estimated distance to the pulsar, which suggests that we will not be able to correct for extrinsic acceleration effects sufficiently, we expect to be unable to employ the $\dot{\omega}$-$\dot{P}_\text{b}$ measurement technique used on PSR~J0737$-$3039 \citep{kwk+16}. However, the likelihood of a significant misalignment angle allows an alternate test using the contribution of Lense-Thirring precession to the rate of change of the projected semi-major axis, given by
\begin{equation}\label{eqn: xdot}
\dot{x}_\text{LT} = x\cot i\left(\frac{\text{d}i}{\text{d}t}\right)_\text{LT},
\end{equation}
where $(\text{d}i/\text{d}t)_\text{LT}$ is given by Eq.~3.27 in \cite{dt92}. Adopting a typically-assumed pulsar moment of inertia of $I = 1.2 \times 10^{45}\,\text{g\,cm}^2$ \citep{ls05} and neglecting the likely slower-spinning companion NS, we calculate that $|\dot{x}_\text{LT}|$ could be as large as $1.9 \times 10^{-14}\,\text{lt-s\,s}^{-1}$. Based upon a continuation of our current timing setup, and assuming additional MeerKAT observations consisting of one orbit/month commencing in mid-2018, we predict a future measurement of $\dot{x}_\text{LT}$ to within $3\,\sigma$ in $\sim8-9$ years. The corresponding (geodetic) spin precession of the pulsar (expected to be $\sim3.1\,^\circ\,\text{yr}^{-1}$) is expected to cause changes in the pulse profile and polarisation, which may allow a determination of the pulsar's spin orientation \citep[see e.g.][]{kra98}.

PSR~J1757$-$1854 is also an ideal system for measuring the PK parameter $\delta_\theta$, which describes the relativistic deformation of the elliptical orbit \citep{dd85}. To date, $\delta_\theta$  has been measured only in PSR~B1913$+$16 \citep{wh16} and PSR~J0737$-$3039 (Kramer et al., in prep.), in both cases with low significance.  As described in \cite{dd86}, the timing residual contribution of $\delta_\theta$  can be characterised by
\begin{equation}\label{eqn: dtheta timing}
\Delta_{\delta_\theta} \simeq 
  -\delta_\theta\,\frac{e^2}{\sqrt{1-e^{2}}}\,x\cos\omega\sin u,
\end{equation}
where $u$ is the eccentric anomaly. The strong dependence of $\Delta_{\delta_\theta}$ on $e$ implies that PSR~J1757$-$1854 (along with other high-$e$ relativistic binaries such as PSRs~B1913$+$16 and B2127$+$11C) will show the strongest timing effects due to $\delta_\theta$. However, Equation \ref{eqn: dtheta timing} also indicates that a measurement of $\delta_\theta$ requires a significant change in $\omega$ in order to separate the residual effect of $\delta_\theta$ from that of $\gamma$ (for which $\Delta_\gamma = \gamma\sin u$). With its high $\dot{\omega} \simeq 10.37\,^\circ\,\text{yr}^{-1}$, PSR~J1757$-$1854 is therefore uniquely positioned to allow for a future measurement of $\delta_\theta$ within a comparatively-short timeframe. Based on the same timing considerations as outlined for $\dot{x}_\text{LT}$, we predict a 3-$\sigma$ measurement of $\delta_\theta$ will be possible within $\sim 7-8$ years.

Finally, as noted in Section \ref{sec:params}, PSR~J1757$-$1854 has the largest observed $\dot{P}_{\rm b}$ and the largest shift in periastron time due to GW emission of any known binary pulsar. This promises a further high-precision test of GR's quadrupole formula for GW damping, as conducted previously with PSR~B1913$+$16 \citep{wh16} and PSR~J0737$-$3039 \citep{ksm06x,kra16}. Timing simulations indicate a test precision of $<1\%$ in only $\sim5$ years. Taking the Galactic potential of \cite{mcm17} and our previous  systemic velocity estimates, we anticipate that the uncertainties on the individual estimates of the distance to PSR~J1757$-$1854 \citep[see][]{NE2001a, YMW16} will limit this test to within a few tenths of a percent.

\section*{Acknowledgements} 

The Parkes Observatory is part of the Australia Telescope National Facility which is funded by the Australian Government for operation as a National Facility managed by CSIRO. The Green Bank Observatory is a facility of the National Science Foundation operated under cooperative agreement by Associated Universities, Inc. Pulsar research at the Jodrell Bank Centre for Astrophysics and the observations using the Lovell Telescope are supported by a consolidated grant from the STFC in the UK. This work is also based on observations with the 100-m telescope of the Max-Planck-Institut f{\"u}r Radioastronomie at Effelsberg. This work is supported by the ARC Centres of Excellence CE110001020 (CAASTRO) and CE170100004 (OzGrav). Survey processing was conducted in association with CAASTRO at the Australian National Computational Infrastructure high-performance computing centre at the Australian National University. The authors wish to thank Marina Berezina, Eleni Graikou, Alex Kraus and Laura Spitler for their assistance with observations at Effelsberg, along with Natalia Lewandowska and Ryan Lynch for their assistance with observations at the GBT. We also thank West Virginia University for its financial support of GBT operations, which enabled some of the observations for this project. AC acknowledges the support of both the International Max Planck Research School for Astronomy and Astrophysics at the Universities of Bonn and Cologne, and the Bonn-Cologne Graduate School of Physics and Astronomy. PCCF and AR gratefully acknowledge financial support by the European Research Council for the ERC Starting grant BEACON under contract No. 279702, and continued support from the Max Planck Society. MK, RK and RPE gratefully acknowledge support from ERC Synergy Grant `BlackHoleCam' Grant Agreement Number 610058. MAM was supported by NSF award number AST-1517003. DRL was supported by NSF award number OIA-1458952.




\bibliographystyle{mnras}
\bibliography{J1757_paper} 

\bsp	
\label{lastpage}
\end{document}